\documentclass[preprint,number,12pt]{elsarticle}
\usepackage{threeparttable}  

\usepackage{amssymb}

 \usepackage{lineno}

\makeatletter

\newcommand{\Rmnum}[1]{\expandafter\@slowromancap\romannumeral #1@}
\makeatother

\begin{document}
\begin{frontmatter}
\title{Maxent in Experimental 2$\times$2 Population Games}
\author{Bin Xu$^{a,b}$}
\author{Zhijian Wang$^{b}$\footnote{Corresponding author. Tel.: +86 13905815529; fax: +86 571 87951328.
E-mail addresses: wangzj@zju.edu.cn (Z.J. Wang)}}

\address{$^{a}$Public Administration College, Zhejiang Gongshang Univ., Hangzhou, 310018, China}
\address{$^{b}$Experimental Social Science Laboratory, Zhejiang Univ., Hangzhou, 310058, China}

\begin{abstract}  
In mixed strategy 2$\times$2 population
games, the realization of maximum entropy (Maxent)
is of the theoretical expectation. We evaluate this theoretical prediction in the experimental economics game data.
The data includes 12 treatments and 108 experimental sessions in which
the random match human subjects pairs make simultaneous strategy moves repeated 200 rounds.
Main results are (1) We confirm that experimental entropy value fit the prediction from Maxent well; and
 (2) In small proportion samples, distributions are deviated from Maxent expectations; interesting is that, the deviated patterns are significant more concentrated.
  These experimental finding could enhance the understanding of social game behavior with the natural science rule --- Maxent.
\end{abstract}
\begin{keyword}  
experimental economics  \sep Maxent  \sep mixed strategy Nash equilibrium \sep population game  \sep 2 by 2 game
\end{keyword}





%
%
%
%
%
%

\end{frontmatter}


 \newpage
 \tableofcontents

\newpage
\begin{linenumbers}
\section{Introduction}

The principle of Maxent is introduced by Jaynes~\cite{Jaynes1957}. By capture the symmetry in stochastic behaviors,  Maxent methodology can provide rich information basing on very limit information~\cite{Golan2008,Wolpert2012}.
Maxent has gained its wide applications in natural science and engineering.
In social science, Maxent approach has also gained its applications, e.g., to
prove the existence of market equilibrium~\cite{Toda2010,Barde2012},   to investigate wealth and income distribution~\cite{Castaldi2007,WuMaxentIncome2003},  to explore firm growth rates~\cite{Alfarano2008} and to build fundamental behavior model~\cite{Wolpert2012}. Theoretical interpreting or modeling of the distributions of social outcomes with Maxent is growing.


Considering the importance of Maxent in social interaction systems,  experimental investigation this fundamental rule is necessary~\cite{Falk2009}.
Only quite recently, entropy is firstly measured out in experiments to evaluate social outcomes by Bednar et.al.~\cite{Yan2011} and Cason et.al.~\cite{Cason2009}; Then, Xu. et.al~\cite{XuetalMaxent2012}
find that fixed-paired two-person constant-sum 2$\times$2 games obey Maxent~\cite{XuetalMaxent2012}. These are the only experimental works, to the best of our knowledge, relates to entropy or Maxent till now.

Comparing with fixed-paired games, population games are more general models for social behavior~\cite{Sandholm2011}. 
Investigating Maxent in population game is expected
because:
(1) Maxent 
 lacks of least experimental population games supporting;
 (2) The experimental precision of Maxent in fundamental experiments in unclear. Robustness experimental results could be the references for future investigations relates to Maxent --- a hub in interdisciplinary.

The main aim of this paper is testing Maxent in experimental population 2$\times$2 games. Randomizing and independence in mixed strategy systems is game theoretician expectation~\cite{VonNeumann1944}, meanwhile, randomizing and independence will lead to the realization of Maxent~\cite{Jaynes2003}. In an ideal (explanations see Section~\ref{experimentdata}) mixed strategy experimental data set from ref.~\cite{selten2008}, we find that (1)  experimental entropy values fit the predictions from Maxent well; and
 (2) whenever the experimental distributions are deviated from  Maxent predictions, the deviated distribution patterns are concentrated significant. Quantitatively, this report provides the precision of Maxent in experimental economics systems.

This report is organized as following. In next section the fundamental 2 by 2 games experiments are reviewed, meanwhile, the reasons for the choosing of this data set are given. In 3$rd$ session, the experimentrics for entropy and distribution, calculation for Maxent expectations and approximate criterion are explained. Results are reported in 4$th$ section. Then, the experimetrics and the implication of our finding on Maxent are discussed briefly and conclusion last. 



\section{Experiments, Data and Presentation}

\subsection{Experiments and Data}\label{experimentdata}

Mixed strategy 2$\times$2 game is one of the most fundamental and the simplest game used widely from student textbooks to frontier research in interdisciplinary. To investigate the performance of Maxent in human subjects social interaction systems, using this kind of experimental game data to is natural. The data we use in this report comes from Selten and Chmura's experiments~\cite{selten2008}.

The experiments contain 12 games, 6 constant
sum games, and 6 nonconstant sum games. Games were run with 12 independent subject groups for each constant sum game and 6 independent subject groups for each nonconstant sum game. Each independent subject group consisted of four
players $X$ and four players $Y$, interacting anonymously over 200 periods with random matching. There were 864 student subjects of the University of Bonn altogether used in these experiments. All of the game are of unique mixed strategy Nash equilibrium systems. In a session with 200 round, the distribution is suggested to be stationary.  In experimetrics view, in these experiments, the treatment sample number is 12 at treatment level, and the group sample number is 108 at session (group) level. At each experimental round, the 8 individual records can be combined as a social outcome of this round; So in a session, there are 200 observations in strategy space in which the distribution can be estimated.

\begin{table}[htbp]
\caption{ Payoff Matrix of $2 \times 2$ Games}\label{tabPom}
\centering
\begin{tabular}{l|ll}
  \hline
       & ~~$Y_1$ & ~~$Y_2$ \\
   \hline
 $X_1$ & $a_{11},b_{11}$ & $a_{12},b_{12}$ \\
  $X_2$ & $a_{21},b_{21}$ & $a_{22},b_{22}$ \\
  \hline
\end{tabular}
\end{table}

The reasons to choose this data set are (1) The games being mixed strategy, so randomization and independence social outcomes can be expected. (2) The unique equilibrium, so each sample has its own observation mean; (3) The variety of the experimental parameters, so the results less bias; (4) Sufficient  samples for there are 108 groups samples, so the confidence of the results could be expected; (5) The distribution of mixed strategy Nash equilibrium  in strategy space is uniform, so the samples from the data could also be less bias for evaluating social performance. These characters make this data set to be ideal and the unique for Maxent investigation.

As subjects (Players $X$ or players $Y$) do not change their roles in a game, subjects in each role can be modeled as a population. So, these games can be regarded as two-population games.
Supposing for the first population $X$, the strategy set is $\{X_{1}$, $X_{2}\}$ for each agent;
similarly, in the second population $Y$, $\{Y_{1}$,$Y_{2}\}$. The payoff matrix for the 12 games can be presented mathematically in
Table~\ref{tabPom} in which, in each cell, the numeric in left (right) is the payoff for the agent from the player from $X$ ($Y$). For the 12 treatments, the payoff matrix cells are shown in column 2-5 in Table~\ref{tab:ent12g} referring to Table~\ref{tabPom}.

\subsection{Presentation of Social State}

\begin{figure*}\label{present5x5} \centering
 \includegraphics[width=120pt]{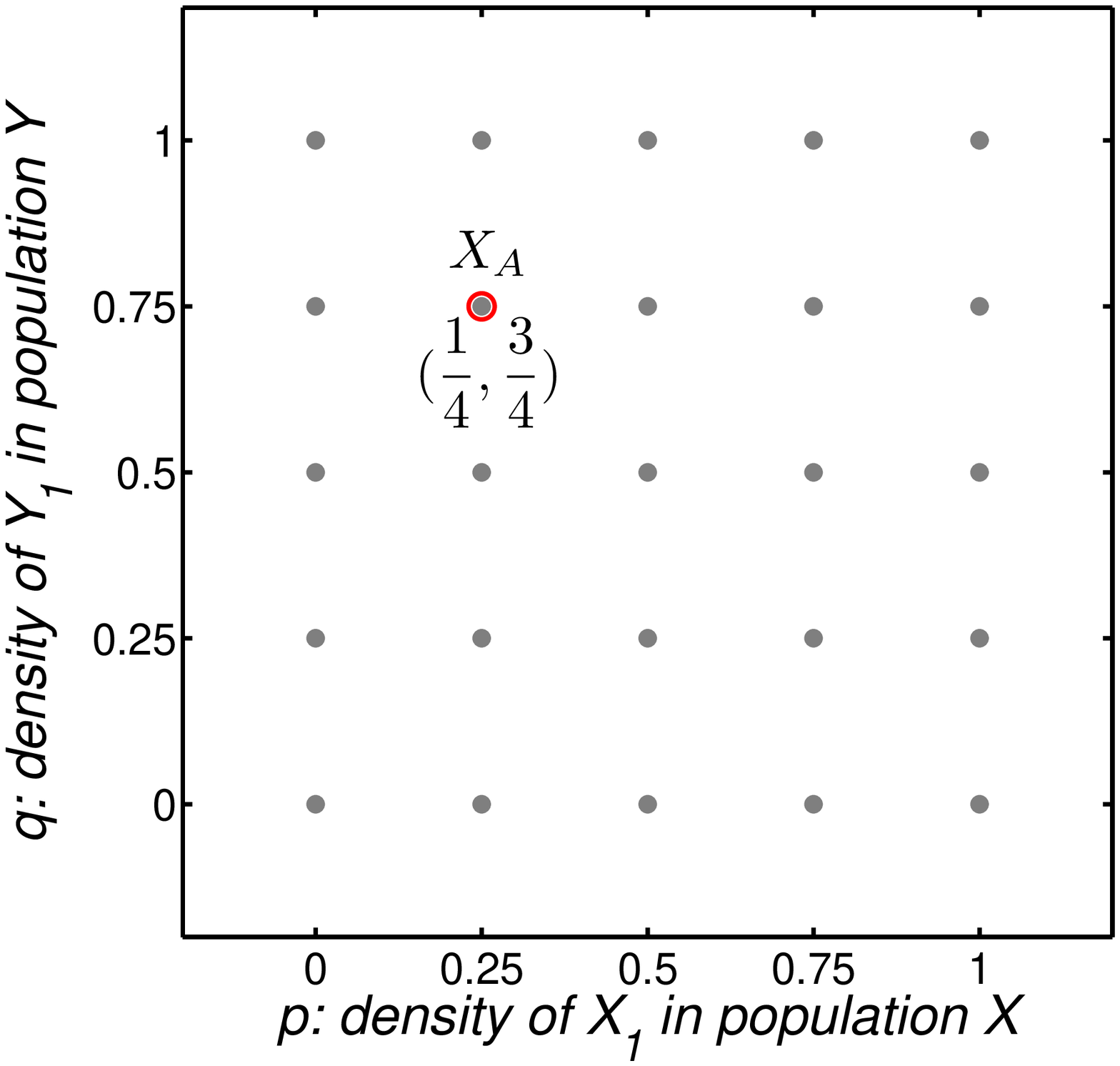}
 \includegraphics[width=120pt]{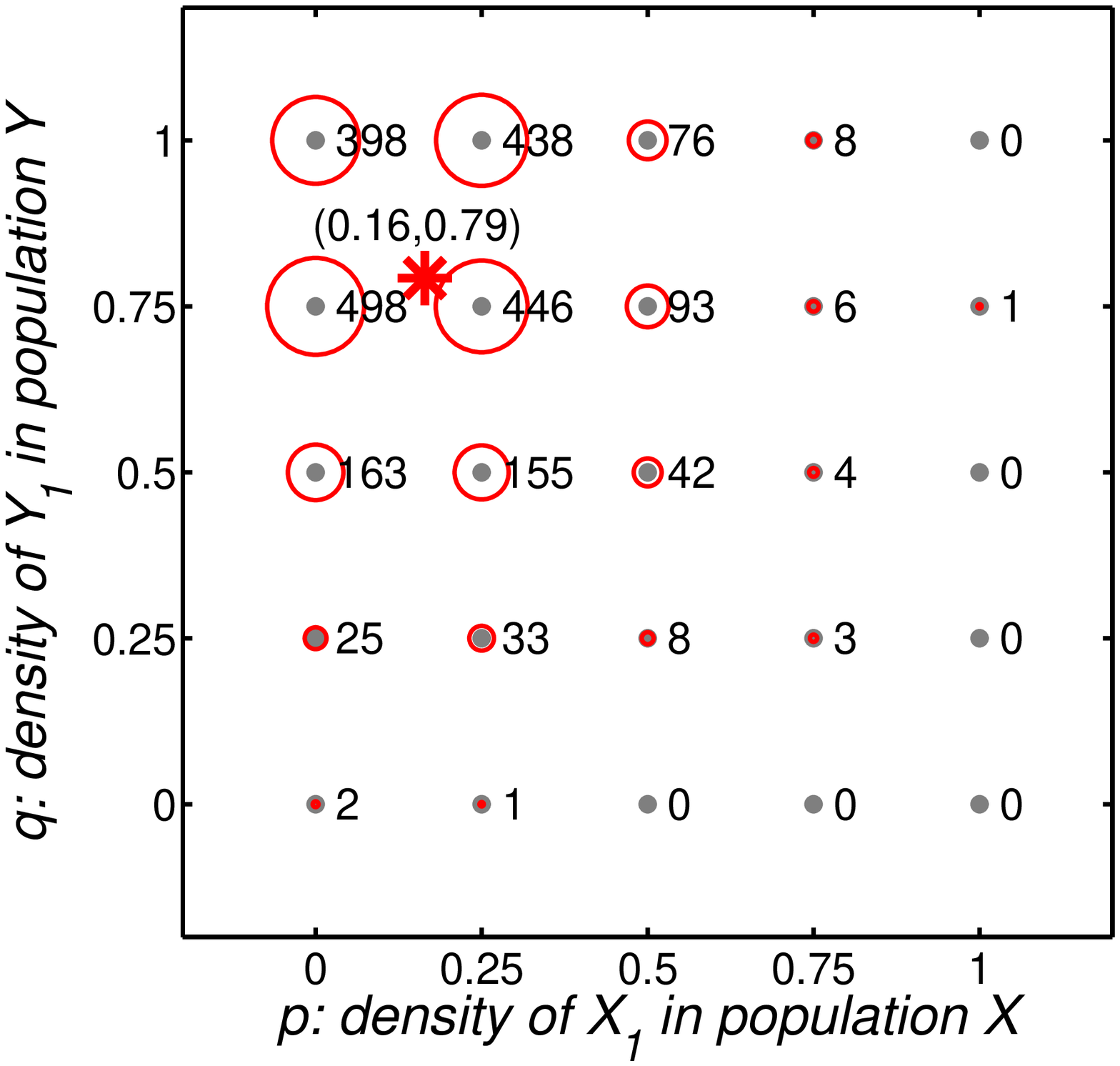}
 \includegraphics[width=120pt]{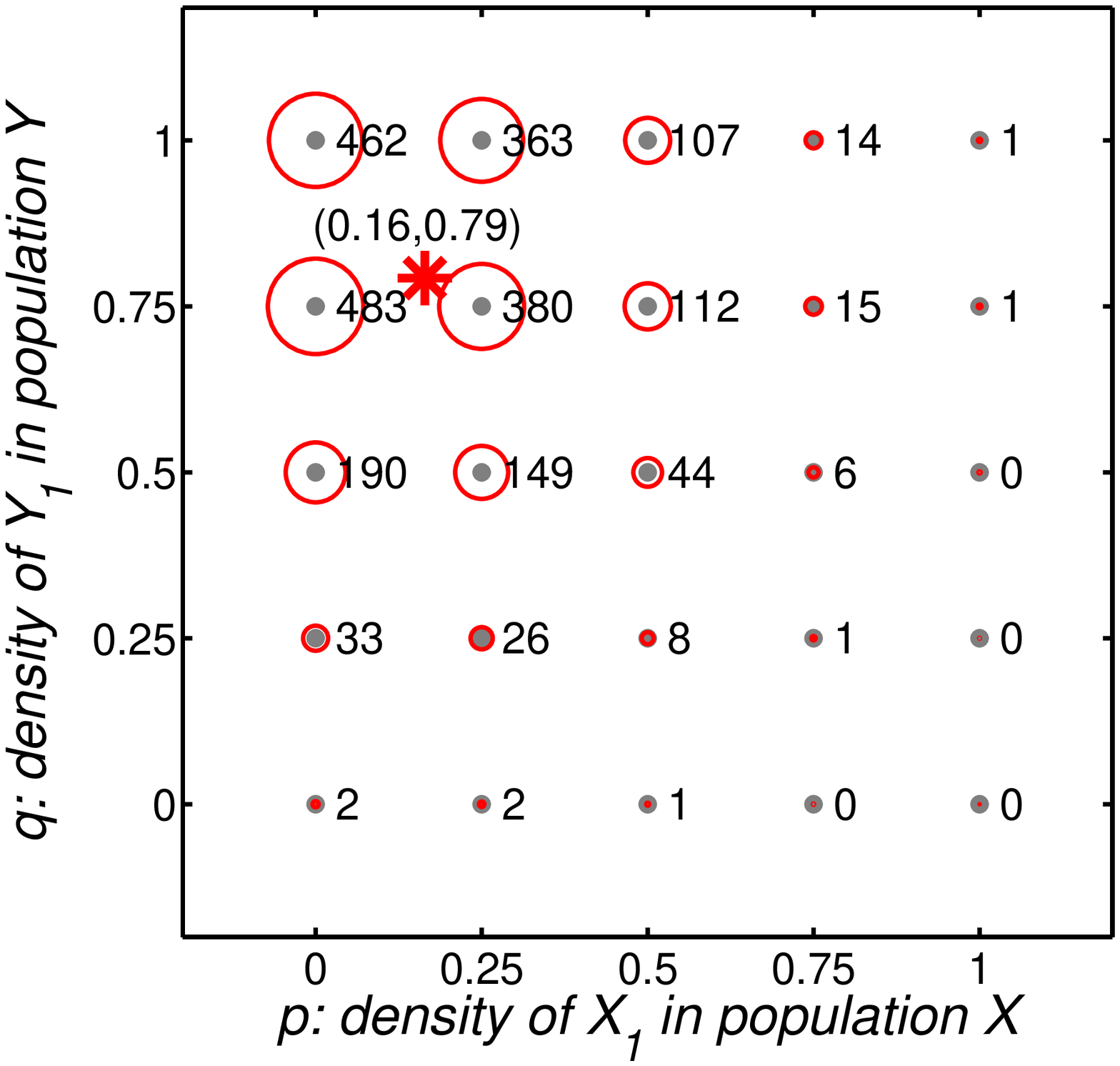}
 \caption{Presentation of the social states lattices and distribution. (Left) A social state which means one in the four agents in $X$-population uses $X_1$, meanwhile  three out of the four agents in $Y$ use $Y_1$. The degree of degenerate is 16 (see the interpretation for Eq.~\ref{eq:disATij}). (Middle) Experimental distribution, data from the Game-3 with 2400 observations, red star is the mean observation ($\bar{O}$)~\cite{selten2008}. (Right) Theoretical distribution from Maxent with the red star ($\bar{O}$) servicing as the given constraints for the Game-3~\cite{selten2008}.  Numerical labels are the observations.}  \end{figure*}

A two-population game's outcomes can be presented in a unit square strategy space~\cite{Friedman1996,Binmore2001} or in a discrete state lattice~\cite{XuWang2011ICCS,Nowak2012}. If there are 4 agents in each of the two population (N=4), an observed instantaneous state should be  $x$:=$(\frac{i}{N},\frac{j}{N})\in \mathbb{X}$, herein $\mathbb{X}$ is the populations strategy state space and  $\mathbb{X}$=$\{0,\frac{1}{4},\frac{2}{4},\frac{3}{4},1\}
 \otimes\{0,\frac{1}{4},\frac{2}{4},\frac{3}{4},1\}$, and $\frac{i}{N}$~($\frac{j}{N}$) is the density of $X_1$~($Y_1$) in $X$~($Y$). Figure~\ref{present5x5} (left) is an illustration of the space and the states. The unit square 5$\times$5 lattice (gray dots) is the space  $\mathbb{X}$. Dots are states. For example, the dot $x_{A}$=($\frac{1}{4},\frac{3}{4}$) is a state. 

 At each round, one observation is gained. So in data, for each treatment 1-6 (7-12) there are 2400 (1200) observations; For each group there are 200 observations. All observations distribute in the 25 dots lattice.

Motivation of using social state presentation is in two aspects. First,
instead of concern the diversity of individual behavior, we concern the performance of social outcomes more~\cite{Friedman1996,Binmore2001,O'Neill1987,Sandholm2011}. And second,
in this presentation,
significant more observations can be harvested from experiment so the regular of social behavior can he evaluated. In sequential strategy presentation, there are $2^8$ states in above experiments; meanwhile, in social state presentation, it is 25 states. 



\section{Measurement and Criterion}\label{secMeasurementCri}

 \subsection{Measurement}
Experimental distribution ($O_{ij}$) at state $(\frac{i}{N},\frac{j}{N})$ can be obtained in data. Every experimental round, an observation is gained in one state in the 5$\times$5 lattice. The distribution $O_{ij}$ can be obtained by pooling up all observations from the experimental rounds. For example, in Fig.~\ref{present5x5}(middle),  state-($\frac{1}{4},\frac{3}{4}$) is labeled as 466 means $O_{(\frac{1}{4},\frac{3}{4})}$=446.  In normalized form, density of state $\rho_{ij}$ can be calculated as $\rho_{ij}:=O_{ij}/T$ in which $T$ is the total experimental rounds.

 The entropy ($S$) can be calculated with a given distribution. Recall that, the original metric for entropy is
 \begin{equation}
 S =-\sum_{k} \rho(k) \log_{\gamma} \rho(k) ,
 \end{equation}
in which $k$ is an element in the individual sequential states set whose independent element number is $2^8$; and $\gamma$ is the potential information in one fundamental observation~\cite{Shannon1948}. This point differs from ref.~\cite{Yan2011,Cason2009} and should be discussed last paragraph in Section~\ref{discussTech}.
In social state presentation, the entropy  can be expressed   as
 \begin{equation}\label{eq:entropygroastgained entropy}
   S=-\sum_{ ij}[\rho_{ij} \log_{\gamma} \rho_{ij}-\rho_{ij}\log_{\gamma}D_{ij}];
\end{equation}
here, $D_{ij}$ is the degeneracy of state $(\frac{i}{N},\frac{j}{N})$ and its arthogram sees the interpreting of Eq~\ref{eq:disATij}. 

Experimental mean observation ($\bar{O}$) can be calculated from ($O_{ij}$) as
\begin{equation}\label{Omean}
    \bar{O} =\sum_{ij} \rho_{ij} (\frac{i}{N}\hat{e}_p + \frac{j}{N}\hat{e}_q)
\end{equation}
in which $\rho_{ij}$ is the density of state at $(\frac{i}{N},\frac{j}{N})$  and $(\hat{e}_p,\hat{e}_q)$ are the two unit vectors. $\bar{O}$ is a $2$-dimensional vector and could be presented as $(\bar{O}_p, \bar{O}_q)$ who server as constraint conditions, which is most often used~\cite{MaxentAvg2012}, for Maxent estimation.

  \subsection{Theoretical expectation}

 Maxent \emph{requires} that the noise process be assigned a  \emph{binomial  distribution} in discrete case, independent and identically distributed ---  this describes our state of knowledge about the noise~\cite{Jaynes2003}.
The theoretical expected distribution ($E_{ij}$) at the state $(\frac{i}{N},\frac{j}{N})$, with the experimental mean observation ($\bar{O}$),  can be calculated as~\cite{Jaynes2003}
\begin{equation}\label{eq:disATij}
    E_{ij} = C_{N}^{i}\!C_{N}^{j} \bar{O}_p^i (1-\bar{O}_p)^{N-i} \bar{O}_q^j (1-\bar{O}_q)^{N-j},
\end{equation}
in which $N$ is the population size ($N$=4 in our case),  $C_{N}^{k}$ is binomial coefficient
 and $C_{N}^{i}\!C_{N}^{j}$ is the degree of degenerate (denoted as $D_{ij}$ above) of the state; For example, at $x_A$ in Fig.~\ref{present5x5}(left), $D_{(\frac{1}{4},\frac{3}{4})}$ is $C_{4}^{1}\!C_{4}^{3}$=16.
Now, with Eq.~\ref{eq:entropygroastgained entropy} and Eq.~\ref{eq:disATij}, theoretical expected entropy ($S_t$) can be obtained.
%

 \subsection{Approximate estimators of the goodness of fit of Maxent}

On entropy view, the approximate criterion to test the goodness of Jaynes Maxent prediction is the entropy concentration theorem (ECT). The theorem   states that out of all distributions that satisfy the observed data (moments),
a significantly large portion of these distributions are concentrated sufficiently close to the one of maximum entropy~\cite{Jaynes1982,Golan2008}.
The measured entropy from experiment $S_{e}$ can be expressed as~\cite{Jaynes1982}
\begin{equation}\label{eq:deltaSr}
S_{t} - \triangle S \leq S_{e}   \leq  S_{t} ,
\label{eq:entropyError}
\end{equation}
in which $S_{t}$ is evaluated from Maxent; Meanwhile, $\triangle S$ can be calculated by 
 \begin{equation}\label{eq:deltaS}
2M \triangle\!S = \chi^2_k (1-F),
\label{eq:entropyError2}
\end{equation}
in which ($1-F$) is the upper tail area (denoted as $p:=1-F$ called as statistical significant index); $k=n-2-1$ is the degrees of freedom in which the 2 is the constrained freedoms due to $\bar{O}$ (the experimental mean observation), $n$=25 is the number of the social states, and then the freedom in the experiments are of 22. From $\chi^2$ table, $\chi^2_{22} (1-0.95)$ is 33.92. So the $\triangle S$ values are of 0.0071, 0.0141 and 0.0848 for the Treatment 1-6 game ($M$=2400) conditions, the Treatment 7-12 game ($M$=1200) conditions and the 108 group ($M$=200) condition, respectively.

On distribution view, the $\chi^2$ goodness of fit test is used to summarize the discrepancy;
 The $\chi^2$ statistic is
\begin{equation}\label{eq:chiSquare}
    \chi^2 = \sum_{i\in S} \frac{(O_{ij}-E_{ij})^2}{E_{ij}};
\end{equation}
Here $O_{ij}$ ($E_{ij}$)  is the experimental (theoretical) distribution and  is given by Eq.~\ref{eq:entropygroastgained entropy} (Eq.~\ref{eq:disATij}).

One point needs to emphasis. Theoretically, both of the criterion above for Maxent are indirectly statistical methods, or saying, both of the criterion have its shortcoming facing experimental economics condition. In the theoretical derivation of $\triangle\!S$, the deviate rate of observe from theoretical expected rate needs to be sufficient small and so high order term can be ignored. However, in the edges of the 25 social states lattice, the experimental observation is rare. Practically, the requirement of the sufficient small can not be satisfied. On the other side,
the $\chi^2$ goodness of fit test is not valid if the expected frequencies are too small. There is no general agreement on the minimum expected frequency allowed, even though values of 5 are often used. However, in the edge of the strategy  lattice, to gain 5 observations is not easy especially for the mean observation closes to the edge of the lattice.
It is a dilemma, if the few observed state are pooled, the good performance of the theoretical prediction on these states is hidden, the degree of freedom is lower.
So we call above two criterions as \emph{approximate estimators} for Maxent in these experimental conditions.

\section{Results}
To test the fit within the  Maxent expectations and the experimental results, the two observable (1) entropy and (2) distribution are used. Results and the supporting materials are reported following.

\subsection{Experimental Entropy and Precision of Maxent prediction}

Experimental entropy values can be calculated from the experimental distribution (Eq.~\ref{eq:entropygroastgained entropy}). Meanwhile, the theoretical entropy can be calculated using the mean observation ($\bar{O}$) and Eq.~\ref{eq:entropygroastgained entropy} with Eq.~\ref{eq:disATij}.

\begin{figure*}\label{EntVal2} \centering
 \includegraphics[width=180pt]{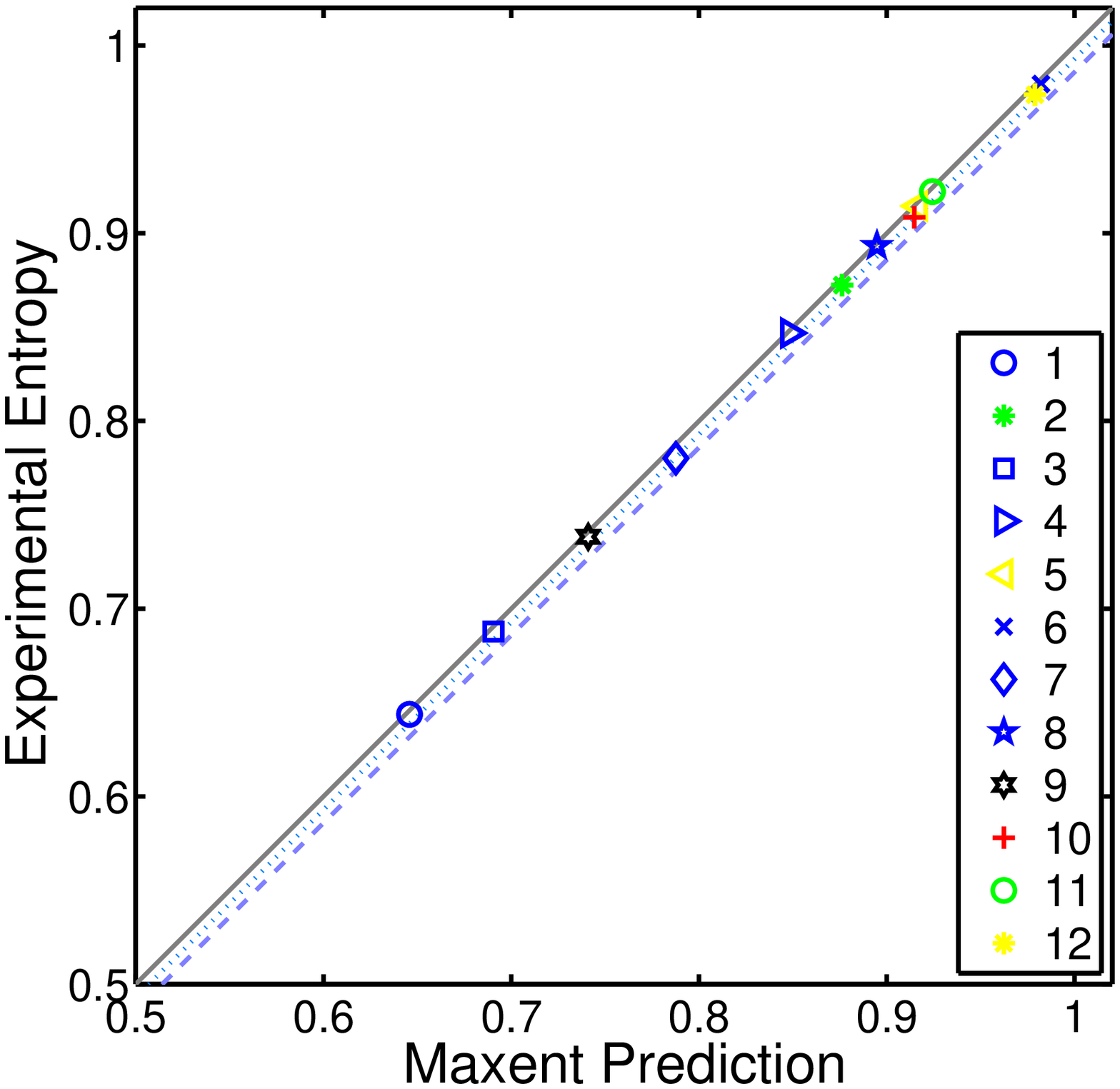}
 \includegraphics[width=180pt]{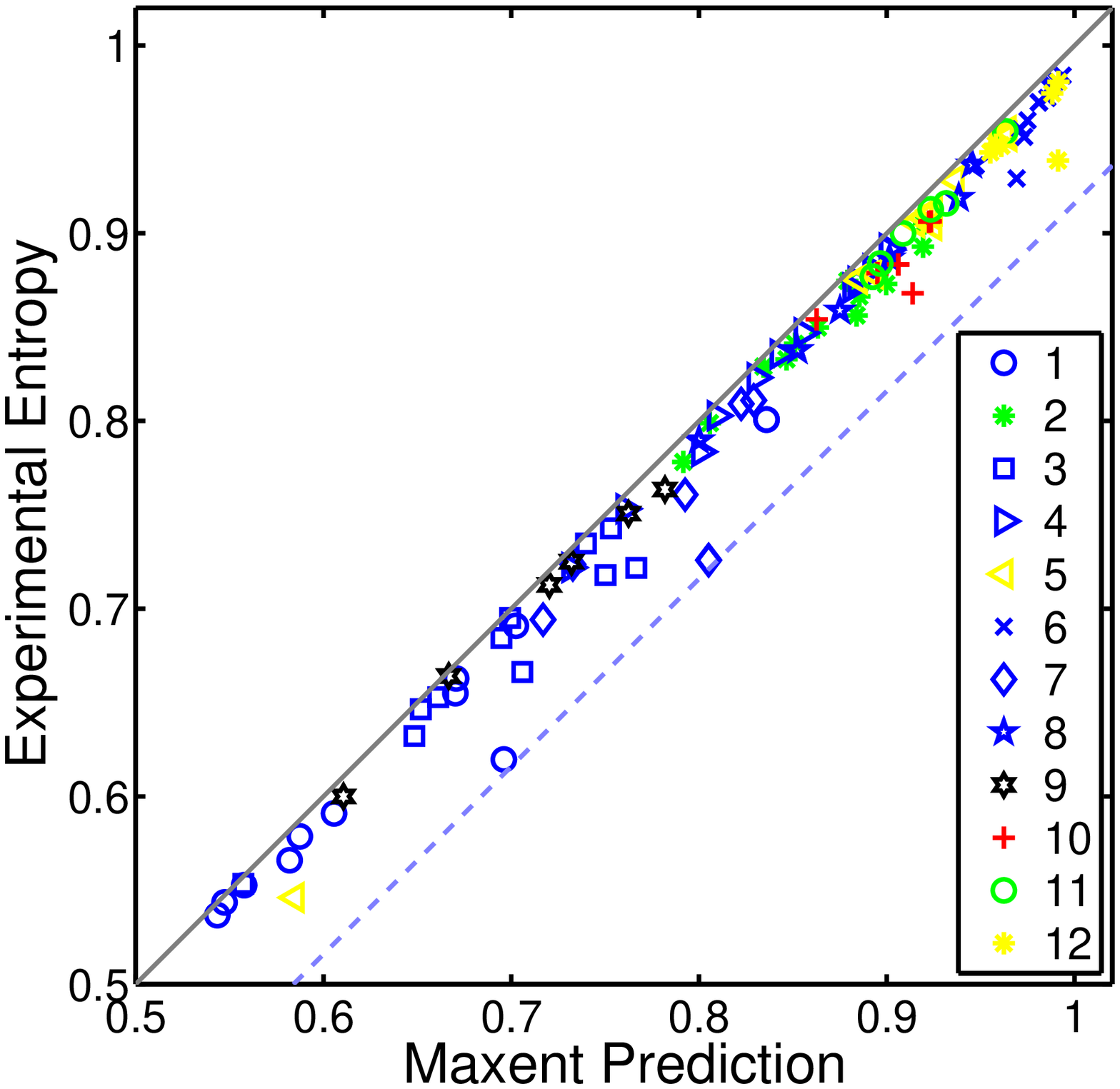}
 \caption{Experimental entropy (in vertical)  vs Maxent theoretical expectation of the entropy (in horizon) in treatment level (left) and in group level (right). Legend  indicate the 12 treatments (from Game 1 to Game 12 referring to ref~\cite{selten2008} respectively). The thin dash indicates the theoretical low-bound  of $S_e$ calculated by Eq.~\ref{eq:deltaS}.
 }  \label{MainFig1s}\end{figure*}

Figural results for eye guide, shown in Figure~\ref{EntVal2},  demonstrates the precision of the Maxent prediction meet the experimental entropy values in treatment level (12 samples, left) and group level (108 samples, right).

Numerical results of the accuracy for the 108 sessions is shown in Table~\ref{tab:EntDiv};
The proportion of the deviation $D_{t-e}$ ($D_{t-e} :=1- \frac{S_e}{S_t}$) are listed along the  treatment and group (in T-G columns). Directly statistical results for the deviation entropy ($D_{t-e}$) are group by treatment is shown in Table~\ref{tab:ent12g}. So, we comes to the conclusion that, prediction of the Maxent suggestion is supported in experimental economics data at accuracy about $0.020\pm 0.002$ level (see the last raw in Table~\ref{tab:ent12g}). In words, in view of entropy values in the samples, the experimental entropy  meets the Maxent prediction well.

\begin{table}[htbp]
\caption{Payoff matrix and the Deviation of $S_e$ from $S_t$  of 12 Treatments}
\label{tab:ent12g}
\begin{tabular}{|c| ccc	c	|| r r	c |}
\hline
T-Gs&	$a_{11}$,	$b_{11}$&	$a_{12}$,	$b_{12}$&	$a_{21}$,	$b_{21}$&	$a_{22}$,	$b_{22}$&	Mean&	S.E.&	 [99$\%$ c.i.]\\
\hline
1,	12&	10,	8&	0,	18&	9,	9&	10,	8&	0.030&	0.009&	0.008,	0.052\\
2,	12&	9,	4&	0,	13&	6,	7&	8,	5&	0.017&	0.003&	0.011,	0.024\\
3,	12&	8,	6&	0,	14&	7,	7&	10,	4&	0.024&	0.006&	0.009,	0.038\\
4,	12&	7,	4&	0,	11&	5,	6&	9,	2&	0.011&	0.001&	0.009,	0.014\\
5,	12&	7,	2&	0,	9&	4,	5&	8,	1&	0.017&	0.005&	0.005,	0.029\\
6,	12&	7,	1&	1,	7&	3,	5&	8,	0&	0.015&	0.003&	0.008,	0.022\\
7,	6&	10,	12&	4,	22&	9,	9&	14,	8&	0.037&	0.013&	0.003,	0.071\\
8,	6&	9,	7&	3,	16&	6,	7&	11,	5&	0.016&	0.002&	0.011,	0.020\\
9,	6&	8,	9&	3,	17&	7,	7&	13,	4&	0.013&	0.003&	0.006,	0.021\\
10,	6&	7,	6&	2,	13&	5,	6&	11,	2&	0.023&	0.006&	0.008,	0.038\\
11,	6&	7,	4&	2,	11&	4,	5&	10,	1&	0.013&	0.002&	0.009,	0.017\\
12,	6&	7,	3&	3,	9&	3,	5&	10,	0&	0.019&	0.007&	0.002,	0.037\\
\hline
 Total&	 &	 &	 &	 &	0.020&	0.002&	0.015,	0.023\\
\hline
  \end{tabular}
\end{table}


A theoretical approximate estimation of the low-bound of experimental entropy ($S_e$) is in Eq.~\ref{eq:deltaS}.
Using $\triangle S$ ($:=S_t - S_e$) from these 108 experimental sessions as samples, the mean deviation ($\triangle S$) from Maxent is 0.0158, the standard error is 0.0012 and the 99$\%$ confident interval (c.i.) is between [0.0125 0.0190]. Calculated from Jaynes ECT (Eq.~\ref{eq:deltaS}), the criterion of up bound of $\triangle S$ = 0.0848 which is large than up bound of the 99$\%$ c.i. significant. This approximate estimation supports above result that the experimental entropy  meets the Maxent prediction well.

\begin{table}[htbp]
\begin{center}
\caption{Deviation ($D_{t-e}$:=1-$S_e / S_t$) of Entropy at Group Level}\label{tab:EntDiv}
\begin{tabular}{|cr|cr|cr|cr|cr|cr|}
\hline
T-G &	$D_{t-e}$ &	T-G &	$D_{t-e}$ &	T-G &	$D_{t-e}$ &	T-G &	$D_{t-e}$ &	T-G &	$D_{t-e}$ &	T-G &	$D_{t-e}$ \\
\hline
1-1&	.042&	2-7&	.03&	4-1&	.01&	5-7&	.066&	7-1&	.098&	10-1&	.025\\
1-2&	.062&	2-8&	.006&	4-2&	.012&	5-8&	.013&	7-2&	.013&	10-2&	.018\\
1-3&	.012&	2-9&	.009&	4-3&	.012&	5-9&	.022&	7-3&	.016&	10-3&	.05\\
1-4&	.009&	2-10&	.012&	4-4&	.01&	5-10&	.017&	7-4&	.022&	10-4&	.018\\
1-5&	.024&	2-11&	.022&	4-5&	.009&	5-11&	.009&	7-5&	.032&	10-5&	.018\\
1-6&	.007&	2-12&	.017&	4-6&	.022&	5-12&	.010&	7-6&	.04&	10-6&	.01\\
1-7&	.012&	3-1&	.033&	4-7&	.009&	6-1&	.009&	8-1&	.017&	11-1&	.018\\
1-8&	.023&	3-2&	.015&	4-8&	.013&	6-2&	.009&	8-2&	.013&	11-2&	.014\\
1-9&	.015&	3-3&	.025&	4-9&	.011&	6-3&	.016&	8-3&	.016&	11-3&	.010\\
1-10&	.110&	3-4&	.007&	4-10&	.008&	6-4&	.014&	8-4&	.009&	11-4&	.009\\
1-11&	.016&	3-5&	.008&	4-11&	.011&	6-5&	.010&	8-5&	.021&	11-5&	.017\\
1-12&	.027&	3-6&	.012&	4-12&	.01&	6-6&	.010&	8-6&	.019&	11-6&	.012\\
2-1&	.016&	3-7&	.006&	5-1&	.011&	6-7&	.012&	9-1&	.01&	12-1&	.015\\
2-2&	.016&	3-8&	.007&	5-2&	.013&	6-8&	.022&	9-2&	.017&	12-2&	.013\\
2-3&	.005&	3-9&	.014&	5-3&	.01&	6-9&	.011&	9-3&	.004&	12-3&	.014\\
2-4&	.031&	3-10&	.056&	5-4&	.013&	6-10&	.041&	9-4&	.011&	12-4&	.011\\
2-5&	.016&	3-11&	.043&	5-5&	.01&	6-11&	.015&	9-5&	.023&	12-5&	.011\\
2-6&	.029&	3-12&	.059&	5-6&	.013&	6-12&	.012&	9-6&	.016&	12-6&	.053\\
\hline
  \end{tabular}
\end{center}
\end{table}

\subsection{Distribution and Deviation Pattern}

\begin{figure*}\label{disdev35} \centering
 \includegraphics[width=250pt]{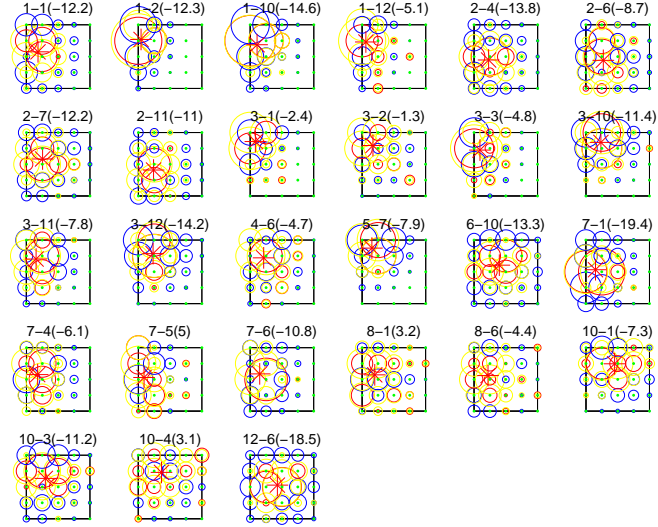}
 \caption{Pattern of the deviation ($O_{ij}-E_{ij}$). Size of the red (blue) cycles indicates the positive (negative) deviation (enlarged 5 times for visible). Size of the yellow cycles indicates the experimental observation $O_{ij}$. In titles, on the three index X-Y(Z), X is Game.ID., Y is Group.ID, And ($Z$), the number in brackets, is results of the sum of deviation.
  }  \label{MainFiag1xz}
 \end{figure*}

Distribution ($O_{ij}$) and the mean observation $\bar{O}$ in an experimental session data can be obtained (Eq.~\ref{Omean}). In the unit square of strategy lattics, for example, the size of the  yellow cycles in figure~\ref{disdev35} indicates the experimental distribution ($O_{ij}$) and the red star indicates the $\bar{O}$ vector for a session; using this $\bar{O}$, theoretical distribution ($E_{ij}$) for a session can be evaluated (Eq.~\ref{eq:disATij}) directly.

An indirect method is used to evaluate the deviation of experimental distribution from Maxent predictions.  
Using Eq.\ref{eq:chiSquare} the $\chi^2$ for each group can be obtained. We use $\chi^2_{22} (1-0.95)$=33.92 as approximation criterion, in total 108 group, there are 27 groups whose $\chi^2$ is larger than 33.92.

 Figure~\ref{disdev35} presents the distribution and deviation of the whole 27 groups. Each of the sub-plot presents a given T-G (treatment-group) group. Results of the distribution ($O_{ij}$) are in yellow; Meanwhile, the positive deviation ($O_{ij}-E_{ij}\!>\!0$) is plotted in red, alternatively, the negative deviation is ($O_{ij}-E_{ij}\!<\!0$) in blue. And the number in brackets ($Z$) is the results of the sum of distance of the deviation of distribution
 in which $Z$:=$\sum_{ij} |x_{ij}-\bar{O}|(E_{ij}-O_{ij})$.
Statistically, $Z$ smaller than 0 ($p<0.001$, $t$-$test$ and $H0$: $Z$=0). This means that, when the Maxent prediction is deviation, on the contrary to more tire-like shape, the binomial bell is, in statistical significant, more sharper.


\section{Discussion and Conclusion}

Maxent has being developed since 1970s to interpret or model  the social outcome distributions~\cite{Toda2010,Barde2012,Castaldi2007,WuMaxentIncome2003,Alfarano2008,Wolpert2012,Golan2008}.
 This report provides, firstly, the laboratory population 2$\times$2 games supporting Maxent (shown in Fig.~\ref{EntVal2}).
The precision of Maxent prediction on entropy values in the fundamental experimental games comes up to 2$\%$.
When experimental distribution is slightly deviates from Maxent prediction, the residual pattern is concentrated significant (shown in Fig.~\ref{disdev35}).
Before interpreting the theoretical and experimental implication, we discuss the experimetric used above.
\subsection{Experimetric Technologies}\label{discussTech}

     In social state presentation, the sequential states, whose social proportions are same, is pooled to one social states. This presentation has been widely used for investigate
     social evolution~\cite{Friedman1996,Binmore2001,O'Neill1987,Sandholm2011,XuWang2011ICCS,Nowak2012}.
     Our investigations are base on this presentation. On the other side, our results have to be strictly constricted on social outcomes, instead of saying that the individual behavior in the games satisfied Maxent.

     Two approximate technologies is used in this report. First is the using of the Jaynes concentration theorem. Indeed we use the $\triangle S$ as a approximation (the dash lines in Figure~\ref{EntVal2} as a eye guide),
      but the results of Maxent well fitted do not relay on the theorem (as explained in last paragraph in Section 3). The results of Maxent well fitted mainly based on the directly measurement results shown in Table~\ref{tab:ent12g} and Table~\ref{tab:EntDiv}. Second is the using of $\chi^2$ goodness of fit approximation to filter out more bias samples.
      This filter out method mainly base on the $\chi^2$ value and larger $\chi^2$ means more deviate from binomial. So, this method do not effect the existence of the sharper-than-binomial pattern in the groups.
So, our main results mainly base on the directly measurements (Table~\ref{tab:ent12g}, Table~\ref{tab:EntDiv} and Fig.~\ref{MainFiag1xz}) instead of relaying on the two approximate statistical technologies. 

     On experimetrics for entropy in Eq.~\ref{eq:entropygroastgained entropy}, the root ($\gamma$) is not 2 as ref.~\cite{Yan2011,Cason2009,XuetalMaxent2012}. We strictly fellow ref.~\cite{Shannon1948} in which the root is the number of all microstate in a device. We set $\gamma$=$2^8$ because for each of the 8 subjects has its own two options, the information in this \emph{social device} is 8-bit. In this way, $S$ has it natural interval $[0, 1]$, and more important, different games can have comparable interval.

\subsection{Implication on Theory}
In mixed strategy game, randomizing using each pure strategy in a certain proportion and playing independent --- these are the fundamental prediction of game theory~\cite{VonNeumann1944}. Meanwhile, in randomization and independence behaviors system, Maxent is a promising~\cite{Jaynes2003} or as a prior~\cite{Wolpert2012}. Using human subjects data, this report makes these two fundamental predictions meet. The finding demonstrates that social behavior could have \emph{common background} as natural science.



 Maxent could help for game theory. Mixed strategy Nash equilibrium is remarkably supported in experiments~\cite{O'Neill1987,Binmore2001}.
However, the mixing is still problem (e.g., Ch.6~\cite{Gintis2009}). Proposal are are developed --- like purification~\cite{Harsanyi1973}, population (people play the same game with different players over time and a mixed strategy is just a distribution of different pure strategies in population), and randomization device (players are playing a pure strategy based on some
randomization device), and so on. Facing these proposal,  a naive directly suggestion is, at least in the 2$\times$2 games, randomization device proposal (for Maxent is mainly supported in data) is dominate; on the contrary, the population proposal is significant less supported.

\subsection{Implication in Experimental Economics}
Entropy measurement has been shown its useful, e.g., ditinguishing the coordination in games~\cite{Cason2009,Yan2011}.
The Maxent methods provides a accuracy way to detect distribution.
For example, in the experiments of testing a fundamental concept of evolutionary game theory~\cite{Nowak2012},
we expect, the deviation of the distribution from Nash (in random and independent hypothesis) could be measured out by Maxent method (using the equations in Section~\ref{secMeasurementCri}). Meanwhile, non-trivial white noise (non-gaussian) phenomena has its extremely interested in theories from multidisciplines~\cite{Stanley1995,Hey2005}. The experimental  Maxent distributions could be an encouragement for the experiments on behavior noise.

We notice that, Maxent methods are not well known in experimental economics.
Our finding suggests that, referring to Maxent prediction,
to test the entropy and distribution of experimental social outcomes is a practical way.

\subsection{Summary}
This report provides firstly experimental evidence of supporting Maxent in
 2$\times$2 population games. The report includes also the precision and the deviation pattern of the experimental social outcomes
 meet Maxent predictions. We wish our experimental finding could strength the bridges from game theory to Maxent and to general science.


\textbf{Acknowledgment:} I am very grateful to R. Selten and K. Binmore
 for helpful suggestions and discussions.  This work is supported by Fundamental Social Science Fund of Zhejiang University.
 \end{linenumbers}

\bibliographystyle{model1a-num-names}












\end{document}